\documentclass[12pt,a4paper]{article}

\usepackage{amsmath}
\usepackage{amsthm}
\usepackage[english]{babel}
\usepackage{graphicx}
\usepackage{amsfonts} 
\usepackage{amssymb} 
\usepackage{hyperref} 
\usepackage{cite}
\usepackage{float}

\usepackage{times}

\newcommand{\oppr}[3]{\ensuremath{\left \langle \left. #1\right.  \right| #2 \left| \left. #3 \right. \right \rangle}}
\newcommand{\ket}[1] {\ensuremath {\left| #1 \right\rangle }}
\newcommand{\bra}[1] {\ensuremath {\left \langle #1 \right|}}
\newcommand{\inpr}[2] {\ensuremath {\left \langle \left. #1 \right| #2 \right\rangle}}

\newcommand{\overroottwo}[1] {\frac{#1}{\sqrt{2}}}
\newcommand{\twomatrix}[4] {\ensuremath{\left[\begin{array}{cc} #1 & #2 \\ #3 & #4\end{array}\right]}}
\newcommand{\twovector}[2] {\ensuremath{\left(\begin{array}{c} #1 \\ #2 \end{array}\right)}}

\newcommand{\fourvector}[4] {\ensuremath{\left(\begin{array}{c} #1 \\ #2 \\ #3 \\ #4 \end{array}\right)}}

\title{Quantum computation from a de Broglie-Bohm perspective} 
\author{Philipp Roser}
\date{\today}

\begin{document}

\maketitle
\begin{center}Department of Physics and Astronomy, Clemson University,\\Clemson, SC 29631, USA \end{center}

\vspace{10mm}	

\begin{abstract}

We consider an example of a quantum algorithm from the point of view of the de Broglie-Bohm formulation of quantum mechanics. For concreteness we look at two particular implementations: one using spin-$\frac{1}{2}$ particles as described by a simple model due to Bell and the other as energy states in an infinite-well potential. We extend the analysis to encompass a complete set of quantum gates. We conclude by discussing the relevance of our investigations for the debate about the origin of the quantum-computational speed-up.

\end{abstract}

\baselineskip = 18pt


\section{Introduction}

Exploiting quantum phenomena for performing computation allows an exponential speed-up in particular types of computational tasks, such as the factorization of large numbers (Shor, 1997, \cite{Shor1997}). While the field of quantum computation has seen dramatic advances in the last two decades, no consensus has been reached regarding the question of exactly which aspects of quantum physics allow the speed-up relative to computations performed by classical means. Numerous proposals have been made, including entanglement (Josza, in Huggett, 1998, \cite{Huggett1998}), the wave-like behaviour of quantum systems (Mermin, 2007, \cite{Mermin2007}) and the existence of ``many worlds'' or ``parallel universes'' (Deutsch, 1985, 1997 \cite{Deutsch1985, Deutsch1997} and in Saunders et al., 2010, \cite{MWT2010}).

We consider it instructive to look at quantum computing from the perspective of another realist theory, de Broglie-Bohm Pilot-Wave Theory (PWT). Whether or not Many-Worlds Theory (MWT) has the explanatory power that some of its proponents say it has with regards to quantum computing, and whether or not it is the only ``interpretation'' of quantum mechanics with that power, is a debate that we will not engage in here. Either way, a fair assessment of the issue relies on a full understanding of quantum computation from the point of view of \emph{any} contending theory. Remarkably, while PWT has received growing interest during the last twenty years or so, the present author has been unable to find an analysis of even a simple quantum algorithm in the context of the theory.

We hope to remedy this situation here by examining the simplest non-trivial example of a quantum algorithm, namely the so-called Deutsch algorithm.\footnote{This is the single-qubit case of the more general ``Deutsch-Josza'' algorithm.} In order to be able to compute de Broglie-Bohm trajectories we will have to make a choice concerning a particular physical model for the qubits. In this paper we examine two such models, spin-$\frac{1}{2}$ particles as described by Bell's simple spin model (Bell, 1987, \cite{Bell1987}), followed by a representation of $\ket{0}$ and $\ket{1}$ as the two lowest energy levels of an infinite square well. We then extend this analysis to a complete set of gates, i.e.\ a set of gates that is sufficient to implement any unitary evolution for an arbitrary number of qubits. Thus the results of this paper show explicitly how PWT can describe elementary quantum computations. Some possible foundational implications are sketched in the conclusion.


\section{Review of the de Broglie-Bohm formulation}

The de Broglie-Bohm pilot-wave formulation of quantum mechanics is a non-local, deterministic hidden-variable theory. Here we will focus on the non-relativistic case, since this will be relevant for our analysis of the quantum algorithms below. Its ontology contains a wavefunction $\psi$, evolving in configuration space according to the Schr\"odinger equation in the position basis as familiar from the standard Copenhagen interpretation,
\begin{equation*}
 i\frac{d}{dt}\psi(\vec{x},t) = H\psi(\vec{x},t)
\end{equation*}
(using $\hbar = 1$), and an actual configuration (the ``particle'' or ``corpuscle'') forming a trajectory in configuration space according to the ``guidance'' equation
\begin{equation*}
 \frac{d}{dt}\vec{x} = \frac{\vec{j}(\vec{x},t)}{|\psi(\vec{x},t)|^2}
\end{equation*}
where $\vec{j}(\vec{x},t)$ is the standard probability current, implicitly defined by
\begin{equation*}
 \frac{d}{dt}|\psi(\vec{x},t)|^2 + \nabla \cdot \vec{j}(\vec{x},t) = 0.
\end{equation*}
For an $n$-particle system with a Hamiltonian 
\begin{equation*}
 H = \sum\limits_{i=0}^n -\frac{1}{2m_i}\nabla_i^2 + V(\vec{x},t),
\end{equation*}
where $m_i$ denotes the mass of the $i$th particle and $\nabla_i$ the derivative with respect to its spatial degrees of freedom, the guidance equation may be simplified to
\begin{equation*}
 \frac{d}{dt}\vec{x}_i = \frac{1}{m_i}\nabla S(\vec{x},t) = \frac{1}{m_i}\mathfrak{Im}(\nabla_i \ln \psi(\vec{x},t))
\end{equation*}
with $\vec{x}_i$ denoting the particle's position in the $i$th particle's subspace of the full configuration space. $S(\vec{x},t)$ is the phase of $\psi(\vec{x},t)$.

A measurement in PWT is the interaction between the system and a measurement apparatus according to an effective Hamiltonian\footnote{The actual Hamiltonian would likely be considerably more complicated given that most apparatuses are made up of very large numbers of atoms.}
\begin{equation*}
 H_{meas} = -iaQ\frac{\partial}{\partial y},
\end{equation*}
where $y$ is a variable for the degree of freedom of the ``pointer'' or ``dial'' on the apparatus from which we may read off the ``outcome'' of the measurement, $Q$ is a unitary operator associated with the particular measurement and agrees with the corresponding measurement operator in the standard formulation, and $a$ is an appropriate coupling constant. The actually observed outcome then depends on the position of the configuration-space particle (see Holland, 1993, \cite{Holland1993}).

Instead of using the guidance equation we may choose to adopt a second-order equation analogous to the Hamilton-Jacobi equation with an extra term (see Bohm, 1952, \cite{Bohm1952}),
\begin{equation*}
 -\frac{\partial}{\partial t}S(\vec{x},t) = \sum_{i=1}^n\frac{(\nabla_i S(\vec{x},t))^2}{2m_i}+V(\vec{x},t)+U(\vec{x},t),
\end{equation*}
where 
\begin{equation*}
 U(\vec{x},t) = \sum_{i=1}^n-\frac{1}{2m_i}\frac{\nabla_i^2|\psi(\vec{x},t)|}{|\psi(\vec{x},t)|}
\end{equation*}
is known as the ``quantum potential'' due to its role in the equation being analogous to the potential $V$. Note however that $U$, unlike $V$, is calculated from the wavefunction and strongly time-dependent. In this paper, we will use the first-order formulation of PWT.

If many copies of the same system are available, then this ensemble may be described by density function $\rho(\vec{x},t)$ in configuration space. If $\rho(\vec{x},t)=|\psi(\vec{x},t)|^2$, then we say that the ensemble is in ``quantum equilibrium'' (Valentini, 1991a, \cite{Valentini1991}). It is straightforward to show that the observed frequency of outcomes matches the predictions of the Born Rule of standard quantum mechanics, which states that the possible outcomes of a projective measurement associated with operator $Q$ are the operator's eigenvalues and the probability of obtaining the $i$th eigenvalue $\lambda_i$ given the system's initial state $\ket{\psi}$ is $|\inpr{\psi}{i}|^2$ where $\ket{i}$ is the eigenstate of $Q$ associated with eigenvalue $\lambda_i$.

If $\rho(\vec{x},t)\neq|\psi(\vec{x},t)|^2$, then the ensemble is in non-equilibrium and outcomes of measurement show statistical deviation from the Born Rule. For ensembles with certain non-equilibrium density functions faster-than-light signalling and violations of the uncertainty principle are possible (Valentini, 1991b, \cite{Valentini1991}). However, subject to certain conditions the ensemble will approach equilibrium as it evolves and a theorem analogous to the classical coarse-graining H-theorem may be derived, explaining the empirical adequacy of the Born Rule. This paper is primarily concerned with the equilibrium case, although much of our analysis applies to both equilibrium and non-equilibrium.


\section{Deutsch's algorithm and generating Hamiltonians}\label{DeutschHamiltonians}

The simplest known non-trivial quantum algorithm is the so-called Deutsch algorithm. We are given a black box (the ``oracle'') that implements one of the four possible functions $f:\{0,1\}\rightarrow\{0,1\}$. Which of these it implements is unknown to us. Enumerate the possible functions as follows:
\begin{eqnarray*} 
&&f_0(0)=0;\qquad f_0(1)=0\\
&&f_1(0)=0;\qquad f_1(1)=1\\
&&f_2(0)=1;\qquad f_2(1)=0\\
&&f_3(0)=1;\qquad f_3(1)=1
\end{eqnarray*}
We are specifically tasked to determine if $f$ is constant (i.e.\ $f\in\{f_0,f_3\}$) or balanced (i.e.\ $f\in\{f_1,f_2\}$).\footnote{For other tasks, such as to determine $f$ fully or to determine whether or not $f=f_1$, say, quantum mechanics is not known to allow any computational speed-up.} Classically this task requires two uses of the oracle, once with input `0' and once with `1' since neither by itself narrows down the set of possible functions sufficiently. Exploiting quantum mechanics (exactly what aspects of quantum mechanics those are is the debate to which we hope to contribute through this analysis) however only one use of the oracle is required. 

To see how, we first make the mechanics of the oracle more explicit: Two qubits are passed through the oracle, the first of which serves as providing the argument of the function $f$ implemented by the oracle (call this the ``data qubit'', following the terminology of Nielsen \& Chuang (2000,\cite{NielsenChuang2000})) and it does not change state as it passes through. The second (``auxiliary'') qubit carries the output: At the end of the oracle interaction it will be in a state $\ket{q\oplus f(p)}$, where $q,p\in\{0,1\}$ denote the input states of the auxiliary and data qubits respectively and `$\oplus$' is used to denote addition modulo~2 . The action on superposition states follows from linearity.

To achieve the stated task, begin with the qubits in the combined state $\ket{0}_d\ket{1}_a$, where the subscripts $d$ and $a$ identify the data and auxiliary qubits respectively. We will express all two-qubit states in the product basis $\mathfrak{H}_d\otimes\mathfrak{H}_a$, i.e.\ the tensor product of the two two-dimensional single-qubit Hilbert spaces with basis $\{\ket{0},\ket{1}\}$. First apply a Hadamard gate $\mathbf{H}=\frac{1}{\sqrt{2}}\twomatrix{1}{1}{1}{-1}$ to each qubit, followed by the action of the oracle, which may be stated in compact 2x2 form as
\begin{eqnarray}\label{oracleevolution} 
U_f = \twomatrix{\delta_{0,f(0)}\mathbf{1}+\delta_{1,f(0)}X}{\mathbf{0}}
		{\mathbf{0}}{\delta_{0,f(1)}\mathbf{1}+\delta_{1,f(1)}X},
\end{eqnarray}
where $\delta_{i,j}$ is the Kronecker-delta symbol and $X=\twomatrix{0}{1}{1}{0}$ denotes the first Pauli matrix. Finally act with another Hadamard gate on the data qubit and perform a measurement on it in the $\{\ket{0},\ket{1}\}$-basis. It is easy to verify that an outcome of 0 implies that $f$ is constant and an outcome of 1 that $f$ is balanced. For a more detailed description we refer the reader to Nielsen \& Chuang (2000, \cite{NielsenChuang2000}).

As preparation for our analysis we need to find the generating Hamiltonians for the unitary evolutions of the quantum gates that constitute the circuit, namely the Hadamard gates and the two-qubit evolution $U_f$. The Hadamard gate is generated by a Hamiltonian of the form
\begin{equation*}\label{Hadamardgenerator}
 H_{Had}\cdot T_{Had} = \frac{\pi}{2}\left[\mathbf{H}-\mathbf{1}\right],
\end{equation*}
where $T_{Had}$ denotes the duration of the action of $H_{Had}$ and $\mathbf{1}$ is the identity operator, as may be readily verified. Thus $\frac{1}{T_{Had}}$ may be understood as the parameter determining the ``strength'' of the acting Hamiltonian.

The block-diagonal form of eq.\ \ref{oracleevolution} makes the task of finding the generating Hamiltonian for the oracle action straightforward. It is:
\begin{eqnarray}\label{Horacle} H_{f}T_{or} 
  = \twomatrix{\delta_{1,f(0)}\cdot\frac{\pi}{2}(X-\mathbf{1})}{\mathbf{0}}{\mathbf{0}}{\delta_{1,f(1)}\cdot\frac{\pi}{2}(X-\mathbf{1})},
\end{eqnarray}
where analogously to above $T_{or}$ denotes the duration of the oracle's action.


\section{Bell's spin toy model}\label{spinmodel}
	
Let us first analyse the Deutsch algorithm as it might be implemented by a toy model due to Bell (1987, \cite{Bell1987}). In the context of PWT the model was first studied by Valentini (2002, \cite{Valentini2002}). The model represents such a particle as a complex two-vector without spatial dependence. The only spatial degree of freedom is that of the pointer of the spin measurement apparatus. Note that spin is not a ``beable'', that is, there are no degrees of freedom of the configuration space particle associated with spin. 



Deutsch's algorithm uses two qubits and hence we need two spin-$\frac{1}{2}$ particles, in addition to a single measurement device, characterised by a variable $y$, that is required for the final step of the algorithm. Initially the wavefunction of the measurement device is a narrow wave packed $\phi_0(y)$ centred at $y=0$. The total initial wavefunction is then given by
	\begin{eqnarray*} \psi_{mn}(y,t_0) = \phi_0(y)d_m(t_0)a_n(t_0), \end{eqnarray*}
where we define $d_m(t)$ and $a_n(t)$ to be the wavefunctions (i.e.\ two-vectors) of the data and auxiliary qubit respectively and $t_0$ is the point in time when the computation is set into motion. 

Calculating the evolution of the pilot wave is then straightforward. Given that the wavefunctions of the two particles are completely specified by a two-component vector of unit length, the two states of the abstract description provided in the previous section can be directly identified with the two eigenstates of $Z=\twomatrix{1}{0}{0}{-1}$:
	\begin{eqnarray*} &&\ket{0}_d \leftrightarrow d_m = \twovector{1}{0}_m, \qquad \ket{1}_d \leftrightarrow d_m = \twovector{0}{1}_m, \\
	&&\ket{0}_a \leftrightarrow a_n = \twovector{1}{0}_n, \qquad \ket{1}_a \leftrightarrow a_n = \twovector{0}{1}_n.
	\end{eqnarray*}
The evolution induced by the two Hadamard gates is given by
	\begin{eqnarray*} \phi(y)d_m(t_0)a_n(t_0) \rightarrow H_{mp}H_{nq} \phi_0(y)d_p(t_0)a_q(t_0) \end{eqnarray*}
and for input states $d_m = \twovector{1}{0}_m$ and $a_n = \twovector{0}{1}_n$ this corresponds to 
	\begin{eqnarray*} \phi_0(y)\twovector{1}{0}_m\twovector{0}{1}_n \rightarrow \frac{1}{2} \phi_0(y) \twovector{1}{1}_m\twovector{1}{-1}_n, \end{eqnarray*}
which we choose to rewrite using pair labels in a single four-component vector, i.e.\ as
	\begin{eqnarray*} \frac{1}{2} \phi_0(y) \fourvector{1}{-1}{1}{-1}_{(mn)} \end{eqnarray*}
with $m,n\in\{0,1\}$.
	
The action of the oracle is then
	\begin{eqnarray*} \frac{1}{2} \phi_0(y) \fourvector{1}{-1}{1}{-1}_{(mn)} \rightarrow \frac{1}{2}\phi_0(y) U_{f\mbox{ }(mn)(pq)} \fourvector{1}{-1}{1}{-1}_{(pq)}. \end{eqnarray*}
	
Denote the time when the final measurement occurs by $t_{meas}$. The pilot wave at a time $t_{meas}-\epsilon$ just prior to the measurement is then given by
\begin{eqnarray*}
  &&\psi_{mn}(y,t_{meas}-\epsilon)=\phi_0(y) \frac{1}{\sqrt{2}}\fourvector{1}{-1}{0}{0}_{mn}
  \qquad \text{if } f(0)=f(1)\\
  &&\psi_{mn}(y,t_{meas}-\epsilon)=\phi_0(y) \frac{1}{\sqrt{2}}\fourvector{0}{0}{1}{-1}_{mn}	\qquad \text{if } f(0)\neq f(1).
\end{eqnarray*}
We note that up to this point the qubits have not interacted with the measurement apparatus, whose wave function $\phi(y)$ is still unchanged. Hence the ensemble equilibrium distribution is unchanged too. 

Following standard de Broglie-Bohm measurement theory (Holland, 1993, \cite{Holland1993}, Bell, 1987, \cite{Bell1987}, Bohm, 1952, \cite{Bohm1952}), the interaction Hamiltonian for the subsequent measurement in this model is given by 
	\begin{eqnarray*} H_{Bell} = -ig(Z\otimes \mathbf{1})\frac{\partial}{\partial y}, \end{eqnarray*}
yielding for a time immediately after the measurement
\begin{eqnarray*} &&\psi_{mn}(y,t_{meas}+\epsilon) = \psi_{mn}(y-2g\epsilon,t_{meas}-\epsilon)	\qquad \text{if } f(0)=f(1)\\
	&&\psi_{mn}(y,t_{meas}+\epsilon) = \psi_{mn}(y+2g\epsilon,t_{meas}-\epsilon) \qquad \text{if } f(0)\neq f(1).
\end{eqnarray*}
Thus, if the initial wave packet of the pointer is sufficiently narrow, we can unambiguously read off whether $f$ is constant or balanced. We note that this result does not strictly rely on being in quantum equilibrium ($\rho=|\psi|^2$), but any ensemble distribution would suffice that evolves in such a way that the two packets of non-zero density separate with the separation of the pilot-wave packets.
	
If we wish we can also trace the evolution of the pilot wave during the times ``inside'' a gate. For example, for the Hadamard gate we found the Hamiltonian
	\begin{eqnarray*}H_{Had}T = \frac{\pi}{2}[H-\mathbf{1}]. \end{eqnarray*}
Setting $t=0$ when the gate is ``switched on'', we may consider the state of the wave at times $t=aT$ where $a\in[0,1]$. The evolution operator is 
\begin{eqnarray*} U = e^{-iH_{Had}aT} &=& e^{-i\frac{a\pi}{2}(H-\mathbf{1})} 
		= \left[\cos\left(\frac{a\pi}{2}\right)\mathbf{1}
		    -i\sin\left(\frac{a\pi}{2}\right)H\right]e^{i\frac{a\pi}{2}}.
\end{eqnarray*}
For $a=1$ this reduces to the complete Hadamard gate and for $a=0$ to the identity. The wavefunction of the pointer (and hence the ensemble distribution) is naturally unchanged while passing through the gate in this model for spin. As such, investigating the evolution ``within'' a gate is only of marginal interest here, although it plays a greater role in our treatment of the infinite-well model..

We have discussed the evolution of the ensemble (unchanged until the measurement, then moving with the shift of the wave packet of the pointer) but have not yet calculated any actual trajectory. Prior to the final measurement the wavefunction of the pointer and that of the qubits are entirely independent. Hence for this period $(t<t_{meas})$ we can write the total wave function as
	\begin{eqnarray*} \psi(y,t) = \phi(y,t)r(t) \end{eqnarray*}
where $r(t)$ is the normalised four-component state vector of the two qubits and $\phi(y,t)$ is the wavefunction of the pointer, such that $\phi(y,t)=\phi_0(y)$ for $t<t_{meas}$. The Hamiltonian $H_{Gate}$ of any of the gates acts only on $r(t)$ and is expressible as a Hermitian matrix. Hence for the current we find unsurprisingly
	\begin{eqnarray*} j(y,t) = -\frac{\partial}{\partial t}|\psi(y,t)|^2 = 0
	\end{eqnarray*}
and so if we continue to neglect the free evolution of the pilot-wave, the equation of motion for $t<t_{meas}$ is simply
	\begin{eqnarray*} \frac{dy}{dt} = 0. \end{eqnarray*}
When the measurement is carried out, the equation becomes
	\begin{eqnarray*}\frac{dy}{dt} = \pm g. \end{eqnarray*}

So not only the wave packet and density distribution as a whole moves with velocity $g$ in the positive or negative $y$-direction depending on the oracle function $f$, but in fact the configuration particle itself moves with this velocity. The trajectories of the ensemble are exactly parallel. The direction of the trajectories depend on the wavefunction. From this we may wish to conclude that most of the ``work'' of the computation is done by the wavefunction and the actual trajectory has little to do with it.


\section{An infinite well model}\label{wellmodel}
	
We will now consider a model in which qubits are realised by the ground state and first excited state of particles in an infinite well, colloquially also known as particles in a box. In particular, to keep the description as simple as possible we will represent qubits by the states of a 1-dimensional well. Since the Deutsch algorithm requires two qubits, we will either need two such particle-well systems, or a single 2-dimensional one. 
	
Assume that the box has a length $L = 1$ and is described by coordinates $x\in[0,1]$. The ground and first excited state in the position basis are then given by
	\begin{eqnarray*} \inpr{x}{0} = \inpr{x}{\psi_1} = \sqrt{2}\sin(\pi x), \qquad\qquad \inpr{x}{1} = \inpr{x}{\psi_2} = \sqrt{2}\sin(2\pi x). \end{eqnarray*}
The discrepancy of index labelling between the wavefunction $(\psi_1,\psi_2)$ and the state kets $(\ket{0},\ket{1})$ arises from our desire not to deviate from conventions used in the general literature. A general state $\ket{\psi} = a\ket{\psi_1} + b\ket{\psi_2}$ evolves into
\begin{eqnarray*} \ket{\psi(t)} 	 
  &=& e^{-i(E_1+E_2)t/2}(ae^{-i\omega t}\ket{\psi_1}+be^{i\omega t}\ket{\psi_2}) 
\end{eqnarray*}
with $\omega = \frac{1}{2}(E_1-E_2)$, which in abstract vector notation corresponds to a state $\twovector{a}{b}$ evolving according to a Hamiltonian $H_{free}=\omega Z$ and so its time evolution is $\twovector{a}{b}\rightarrow R_z(2\omega t)\twovector{a}{b}$, where $R_z(\theta) = e^{-i\theta Z/2}$, plus an overall phase rotation. This means we can implement a $Z$-gate, for example, by simply waiting for the correct amount of time $\omega t=\pi$.

Recall that any single-qubit gate can be written in the form 
\begin{eqnarray*} U_{Gate} = e^{i\alpha}R_z(\beta)R_x(\gamma)R_z(\delta), \end{eqnarray*}
where $R_x(\theta)$ and $R_z(\chi)$ correspond to rotations in the Bloch sphere by angles $\theta$ and $\chi$ around the x and z-axis respectively and that geometrically any arbitrary rotation may be implemented by three sequential rotations round these axes. $e^{i\alpha}$ fixes the overall phase, which in our example may be implemented using the time evolution of the overall phase factor $e^{-i(E_1+E_2)t/2}$.

Performing rotations $R_x(\theta)$ may be achieved by perturbing the potential (and hence the Hamiltonian) inside the well. In particular, adding the term
	\begin{eqnarray*} \delta V(x) = -\frac{9\pi^2}{16}\left(x-\frac{1}{2}\right) \end{eqnarray*}
(see e.g.\ \cite{NielsenChuang2000} p.\ 280) we find that its matrix elements are given by
	\begin{eqnarray*} \oppr{\psi_n}{\delta V}{\psi_m} = \twomatrix{0}{1}{1}{0}_{mn}, \end{eqnarray*}
which can be easily verified by performing the integration explicitly in the position basis. This Hamiltonian thus generates the required rotations $R_x$ around the x-axis in Bloch space.

The Hadamard gate $\mathbf{H} = \frac{1}{\sqrt{2}}\twomatrix{1}{1}{1}{-1}$, for example, is given by
\begin{eqnarray*} \mathbf{H} = e^{i\frac{\pi}{2}}R_z\left(\frac{\pi}{2}\right)R_x\left(\frac{\pi}{2}\right)R_z\left(\frac{\pi}{2}\right), \end{eqnarray*}
corresponding to a free evolution for a period $\Delta t = \frac{\pi}{4\omega}$, followed by a perturbation of the potential by $\delta V$ for a period $\Delta t = \frac{\pi}{4}$ and finally followed by another period of free evolution. We justify the neglect of the free evolution during the second time interval (the X-rotation) by choosing $\omega$ to be small (e.g.\ by choosing a large particle mass $m$). The overall phase will be relevant in interactions with other qubits, but in the Deutsch algorithm there is only one such interaction (the oracle) and hence the phase may be fixed by preparing the correct states at the right time to achieve the right phase factor for the interaction.

We will now turn our attention to the interaction of the oracle. We will consider specific choices $f$. $f_0$ is trivial to implement (no evolution at all) and $f_3$ is simply a NOT-gate in the auxiliary qubit, which we can implement by perturbing the potential as above for an interval $\Delta t = \frac{\pi}{2}$. In neither of these cases is any interactions between the qubits required. However, for the cases of balanced functions $f_i$, i.e.\ for $f_1$ and $f_2$, the situation is different and an actual two-qubit gate needs to be constructed.
	
We will only consider $f_2$ here. A construction of $f_1$ is easily obtainable by modifying the description here, or by using the $f_2$ evolution exactly, but placing an additional NOT-gate before and after the oracle in the data qubit. We implement the oracle by perturbing the potential by an additional term $U(x,y)$, where $x$ and $y$ denote the position coordinates of the data and auxiliary qubits respectively. We remark that if physically we have a single particle in two dimensions, such a perturbation may be easily visualised. For two particles in one-dimensional wells, this term corresponds to a (possibly quite complicated) coupling between them. Equation \ref{Horacle} for the case $f_2$ reads
\begin{eqnarray}\label{f2Hamiltonian} 
T_{or}H_{f_2\text{ }ij} = \twomatrix{\frac{\pi}{2}(X-\mathbf{1})}{0}{0}{0}_{ij}.
\end{eqnarray}

Choosing for simplicity the time interval during which the gate acts to be $T_{or}=\frac{\pi}{2}$, $U(x,y)$ must have the following matrix elements:
\begin{eqnarray*}
  \bra{\psi_k}_d\bra{\psi_m}_a U(x,y) \ket{\psi_n}_a\ket{\psi_l}_d &=& (X-\mathbf{1})_{mn} 
    \qquad\text{for }k=l=1  \\
  \bra{\psi_k}_d\bra{\psi_m}_a U(x,y) \ket{\psi_n}_a\ket{\psi_l}_d &=& 0 \qquad\qquad\qquad\text{otherwise}
\end{eqnarray*}
with $k.l.m,n\in\{1,2\}$. This may be achieved with a potential $U(x,y,)$ inside the well of the form
\begin{eqnarray*} U(x,y) = (A+B\cos x+Cx\cos x)\left[-\frac{9\pi^2}{16}(y-\frac{1}{2})-1\right], \end{eqnarray*}
where $A,B$ and $C$ are appropriately chosen constants. Appropriate numerical values are $A=\frac{52}{27}$, $B=-\frac{225}{432}\pi^2$ and $C=\frac{225}{216}\pi^2$, although other linear combinations of functions of $x$ are also possible. The matrix elements
$\bra{\psi_p}_d\bra{\psi_m}_a U(x,y) \ket{\psi_n}_a\ket{\psi_q}_d$ 
then satisfy the requirements given above as may once again be verified by explicit calculation in the position basis.

The final measurement is an interaction between the data qubit and the pointer described by a position variable $z$ and whose wavefunction starts as an initial narrow wave packet centered at $z=0$. An appropriate interaction Hamiltonian is given by
\begin{eqnarray*} H_{meas} = a(-i\partial_x)^2(-i\partial_z), \end{eqnarray*}
where $a$ is a coupling constant. From PWT measurement theory we know that as a result the ensemble density shifts by $a\delta t\cdot n^2\pi^2$ for $n=1,2$ corresponding to the two energy states, and $\delta t$ denotes the duration of the interaction between qubit and apparatus. The projection of the trajectories onto the configuration space dimension of the pointer are all exactly parallel. 

We see that we can fully account for the outcome of the computation by only considering the wavefunction up to the point of measurement and then appeal to standard PWT measurement theory. As such, we have also automatically analysed the evolution of the ensemble, at least in quantum equilibrium. A corresponding analysis for quantum non-equilibrium would possibly present interesting future work, although we can already conclude that if $\psi = 0 \rightarrow \rho = 0$ (i.e.\ the density function only has ``support'' where the pilot-wave is non-zero),\footnote{For this not to be the case the ensemble density would have to be non-zero somewhere outside the box. The fact that the phase is undefined in such areas will likely be problematic for a coherent PWT description for such ``extreme'' non-equilibrium.} then the computation works exactly the same nevertheless: The measurement outcome is deterministic in every sense (as follows from the quantum formalism) and only depends on the choice of oracle function $f$, hence \emph{any} trajectory whose corpuscle is at some time inside the configuration space area where $\psi\neq 0$ corresponds to the correct computational output. This conclusion does not apply to other, non-deterministic algorithms.
	
The only task left before us is to calculate trajectories of the configuration space particle (in the $x$-$y$ subspace of configuration space) during the action of the gates. Note that the aspect most crucial for the trajectory is its projection onto the dimension corresponding to the degree of freedom of the pointer of the measurement apparatus, which we have already discussed. We may wish to investigate the trajectories of the qubits themselves for completeness. Since we are dealing with systems of spinless particles whose Hamiltonian is of the form $H_{well}=-\frac{1}{2m}\nabla^2+V$, the guidance equation is the usual $m\dot{\vec{x}} = \nabla S$. We are particularly interested in how the trajectories of the particles differ between the cases of a constant and a balanced oracle function $f$. We recall that the data qubit enters the oracle in a state 
$\ket{+}_d  = \sin(\pi x)+\sin(2\pi x))$. If $f$ is constant it will return to that state when the oracle action stops and if $f$ is balanced its final state will be $\ket{-}_d = \sin(\pi x)-\sin(2\pi x))$.
The corresponding quantum equilibrium ensemble density functions in the $x$-dimension are depicted in figure \ref{densities} below.

\begin{figure}[H]
\centering
  \includegraphics[width=0.4\textwidth]{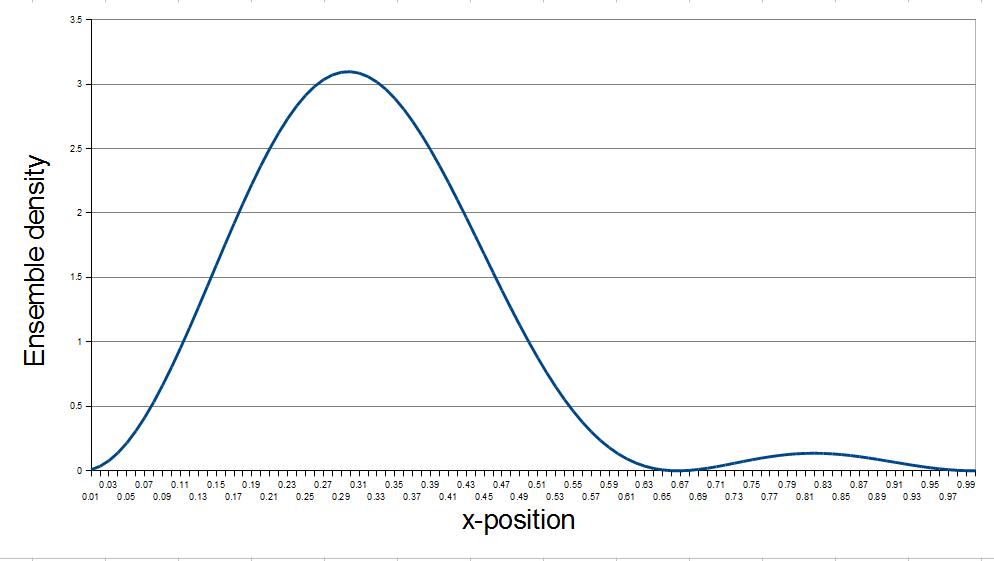}
  \includegraphics[width=0.4\textwidth]{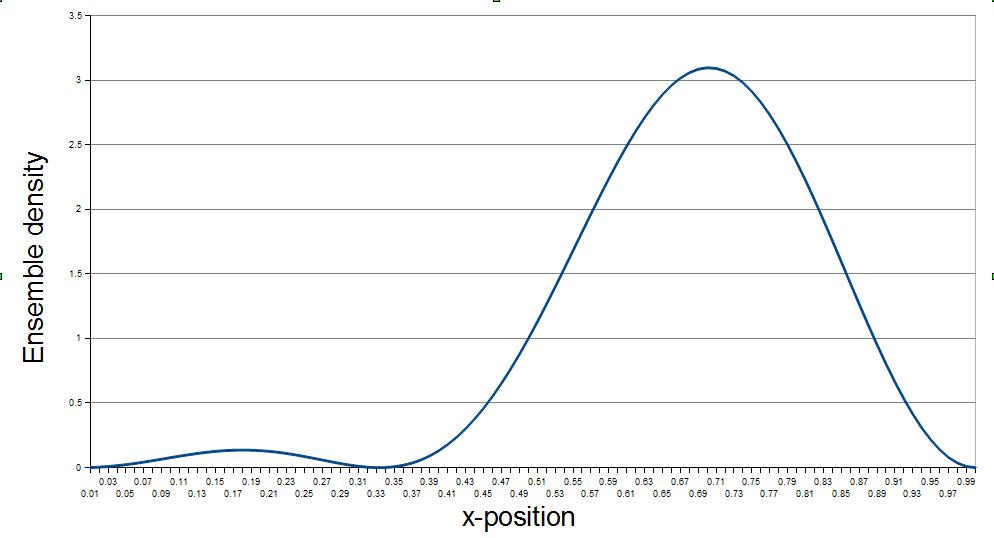}
  \caption{\small Ensemble densities $\rho(x) = |\inpr{x}{+}|^2 = |\sin(\pi x) + \sin(2\pi x)|^2$ and $\rho(x) = |\inpr{x}{-}|^2 = |\sin(\pi x) - \sin(2\pi x)|^2$ for an infinite well of length $L=1$.}
\label{densities}
\end{figure}

This already highlights the key difference between the types of trajectories found in the two cases: In the constant case, the particle ensemble returns to its initial distribution of $x$-positions, in the balanced case it does not. For both a constant and a balanced oracle function, the auxiliary qubit evolves from $\ket{-}_a$ to $\ket{-}_a$ (i.e.\ its initial and final states are the same) and so the ensemble returns to its initial $y$-distribution. 
	
We will now take a glimpse at the trajectories themselves. Since, for pilot waves in non-trivial superpositions of the two eigenfunctions $\sin(\pi x)$ and $\sin(2\pi x)$, the phase $S$ is given by the inverse tangens of a complicated function of $x,y$ and $t$ and the trajectories cannot be found analytically, we analyse these numerically. Here we will only consider the evolution of the configuration particle during the action of the oracle rather than the entire algorithm. We will attempt to compare the two specific cases $f_0$ and $f_2$, i.e.\ one constant and one balanced oracle function $f$. 
	
First, however, consider the free evolution of the qubits in the absence of any gates. The phase $S$ of the states $\ket{0}$ and $\ket{1}$ is just given by $-iE_1t$ and $-iE_2t$ respectively and hence $\nabla S = 0$ in both cases. The particle is at rest. This is not the case for superpositions. The phase of some general state 
\begin{equation*}
\ket{\psi} = ae^{-i\omega t}\sqrt{2}\sin(\pi x)+be^{i\omega t}\sqrt{2}\sin(2\pi x) 
\end{equation*}
 (up to some additional overall phase factor independent of $x$), is given by
\begin{equation*}
S = \arctan\bigg\{ \tan(\omega t)\frac{-a\sin(\pi x)+b\sin(2\pi x)}{a\sin(\pi x)+b\sin(2\pi x)} \bigg\}. 
\end{equation*}
Its gradient $\nabla S$ determines the particle trajectories and results in a one-dimensional oscillatory behaviour in $xy$-space.



Let us now investigate the oracle. For the constant function $f_0$, the corpuscle remains at rest (neglecting its free evolution), from which the ensemble evolution anticipated in figure \ref{densities} follows trivially. Let us contrast this to the evolution generated by the corresponding Hamiltonian, given in equation \ref{f2Hamiltonian}. One result to note immediately is that the particle does not move at all in the $y$-dimension, no matter what its initial values for $x$ and $y$ are. Thus its ensemble density evolution is again recovered naturally. The $x$-evolution is more interesting. The results of the numerical modelling can be seen below in figure \ref{oraclex}. The initial distribution of the particle positions corresponds to (coarse-grained) quantum equilibrium.
	
\begin{figure}[H]
	\centering
		\includegraphics[width=\textwidth]{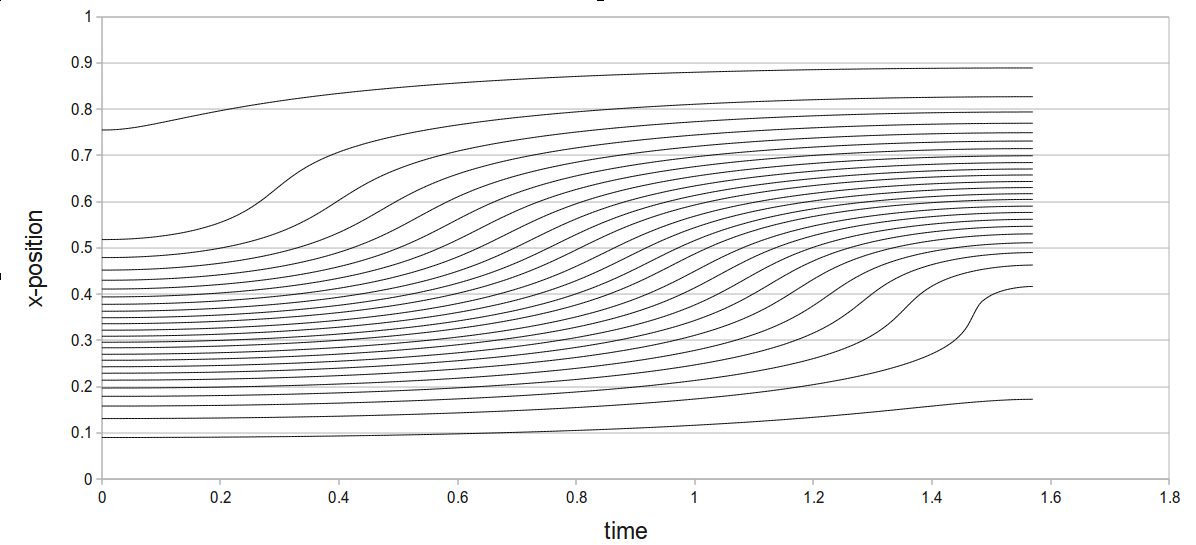}
	\caption{Trajectories in the $x$-dimension for an oracle implementing $f_2$}
	\label{oraclex}
\end{figure}

The trajectories are consistent with our expectation. The ensemble density shifts in $x$-space from its initial to its final distribution in relatively simple paths. Recall that in the case of Bell's spin model spatial motion was only necessary for the pointer in order to account for the computation of the Deutsch algorithm. 

An infinite well is not a realistic realisation fo qubits and parameters for the numerical simulation were chosen arbitrarily. Future work should include more realistic realisations of the qubits where parameters such as the particle mass can be chosen based on physical considerations.


\section{Description of a complete set of gates}

A de~Broglie-Bohm analysis using Bell's model for \emph{any} quantum algorithm is relatively trivial: Abstract $\ket{0}$, $\ket{1}$ states map straightforwardly onto spin states and the only degrees of freedom for the particle are those of the measurement apparatus, which have already been discussed, although in general multiple qubits may be measured.

Let us therefore instead extend our analysis of the infinite-well model. It has been shown (DiVincenzo, 1995, \cite{DiVincenzo1995}) that three gates are together sufficient for any desired unitary evolution for any number of qubits. One possible such set is the Hadamard gate, the $\frac{\pi}{8}$-gate $T=\twomatrix{1}{0}{0}{e^{i\pi/4}}$ (both single-qubit gates) and the (two-qubit) controlled-NOT gate.\footnote{As discussed above, any single qubit gate can be implemented using only $z$ and $x$-rotations, and an overall phase. Hence these together with the CNOT-gate would also form a complete set for constructing unitary gates with any number of qubits.}

The last of these we have already discussed in detail: The function $f_2$ of Deutsch's algorithm corresponds exactly to a controlled-NOT gate (possibly up to a NOT-gate before and after the controlled-NOT, depending on one's preferred definition).

\begin{figure}[H]
	\centering
		\includegraphics[width=\textwidth]{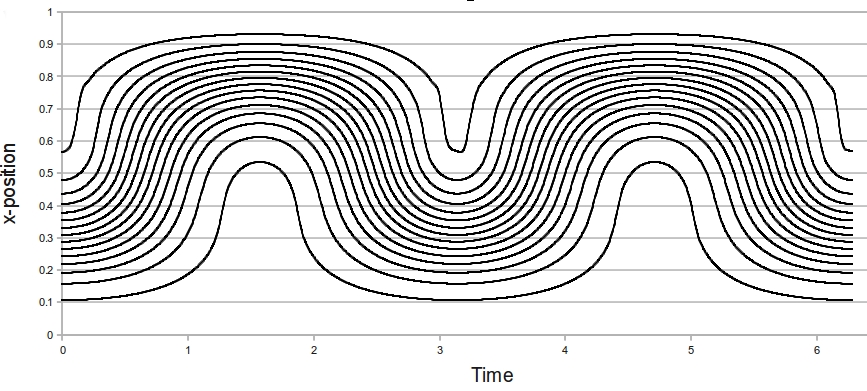}
	\caption{Sample trajectories for the free evolution, i.e.\ with an unperturbed infinite-well potential.}
	\label{free_evolution}
\end{figure}

The $\frac{\pi}{8}$-gate is generated by a Hamiltonian
\begin{equation*}
 H_{\frac{\pi}{8}} = \frac{1}{T_{\frac{\pi}{8}}}\cdot\frac{\pi}{8}(Z-\mathbf{1}),
\end{equation*}
where $T_{\frac{\pi}{8}}$ is the interval during which this Hamiltonian is ``switched on''. An analysis of the $\frac{\pi}{8}$-gate is simple if the initial state is one of the basis states. In that case the evolution is limited to the overall phase and the configuration particle does not move. For other initial states the evolution corresponds just to the free evolution. The phase factors of the two basis states rotate in opposite direction, in addition to an irrelevant change in the overall phase. The resulting behaviour of the particle is an oscillation inside the well that depends on its initial configuration. Numerical analysis for a small equilibrium ensemble of such particles has yielded the trajectories given in fig.\ \ref{free_evolution}.

\begin{figure}[H]
	\centering
		\includegraphics[width=\textwidth]{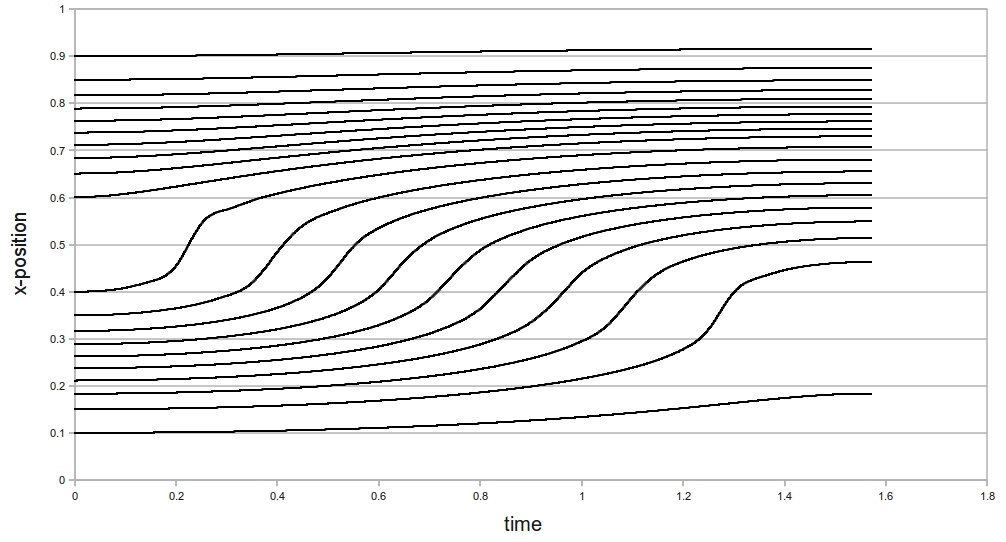}
	\caption{Sample trajectories for the Hadamard gate in the infinite-well model, with initial state $\ket{1}$ corresponding to the well's first excited state.}
	\label{Hadamard_from_1}
\end{figure}

Finally recall the Hadamard gate discussed in section \ref{DeutschHamiltonians}. Consider the concrete evolution example $\ket{1}\rightarrow \mathbf{H}\ket{1}=\overroottwo{1}(\ket{0}-\ket{1})$. The corresponding evolution of the equilibrium distribution in the infinite well therefore takes the form
\begin{equation*}
 \left|\sqrt{2}\sin(2\pi x)\right|^2\quad\longrightarrow\quad 
    \left|\overroottwo{1}\left(\sqrt{2}\sin(\pi x)-\sqrt{2}\sin(2\pi x)\right)\right|^2 .
\end{equation*}
Fig.\ \ref{Hadamard_from_1} shows the evolution of 20 trajectories constituting a coarse-grained equilibrium ensemble.

Of course, here we have only displayed concrete trajectories for chosen initial states. Other initial states could be used, leading to different trajectories, although ultimately no fundamentally new insight would be gained from a more comprehensive analysis in this respect.

While we have provided a description of trajectories for a set of gates that together are sufficient to construct a circuit equivalent to any chosen multi-qubit evolution, it might nonetheless be instructive to look at another algorithm as a whole, bearing in mind that our purpose is not to show that PWT \emph{is} able to account for quantum computation (that already follows from the fact that equilibrium PWT reproduces standard quantum mechanics empirically), but to describe \emph{how} it does so in order to facilitate a discussion of how the relative speed-up of quantum computation can be explained. However, we will leave this to possible future work.

%


\section{Conclusion and Discussion}

We have presented an explicit example of a simple quantum algorithm and a description of a complete set of gates in the context of PWT. While the theorem that equilibrium PWT reproduces all measurement results of standard quantum mechanics exactly already guarantees that quantum computers work in PWT, it is nonetheless instructive to see this done explicitly.

Quantum computation provides insightful and interesting examples of quantum evolutions and measurements. The original motivation for this paper was to understand how such processes could be described in PWT. Having done so, one may ask what implications this work may have for the interpretation of quantum computing and quantum mechanics generally. While in our view fundamental questions concerning quantum theory are not really tied to quantum computing, even if it provides striking examples, some authors have claimed otherwise and it therefore seemed worthwhile to study quantum computing from the perspective of PWT.

What conclusions can we draw? We have seen that the essence of the computation lies with the evolution of $\psi$, even in PWT. This was particularly evident in our analysis in terms of Bell's model of spin. Even so, it should be emphasised that the pilot-wave account is conceptually and physically distinct from that of many-worlds theory, despite some claims to the contrary. In particular, its account of measurement is radically different and outcomes are accounted for in terms of an actual configuration, not a global object such as the wavefunction. $\psi$ is not interpreted as constituting many worlds, but simply as a complex field in configuration space. Everettians may refer to the often cited "many worlds in denial" argument (see Deutsch, 1996, \cite{Deutsch1996}, and Brown \& Wallace, 2007, \cite{BrownWallace2007}) and argue that PWT involuntarily includes the structure of many worlds. However, this argument is contentious (e.g.\ Valentini, in Saunders et al., 2010, \cite{MWT2010}) and furthermore there are, arguably, open issues with the claim concerning the emergence of the many-world structure in the Everettian picture.\footnote{For a discussion of the author's views we refer the reader to Roser (\cite{MSCthesis}).} Either way, it is our hope to contribute to this ongoing debate with the technical analysis provided in this paper.


\section*{Acknowledgements}
The author wishes to thank Antony Valentini for helpful discussions and suggestions.	



\end{document}